\documentclass[showpacs,showkeys,preprintnumbers,amsmath,amssymb]{revtex4}

\usepackage{graphicx}
\usepackage{dcolumn}
\usepackage{bm}

\begin{document}

\preprint{OPCT-July-2008}

\title{High-temperature expansion of the magnetic susceptibility 
        and higher moments of the correlation function 
        for the two-dimensional XY model }

\author{H. Arisue}
\email{arisue@ipc.osaka-pct.ac.jp}
\affiliation{%
Osaka Prefectural College of Technology \\ 
26-12 Saiwai-cho, Neyagawa, Osaka 572, Japan
}%


\date{\today}

\begin{abstract}
  We calculate the high-temperature series of the magnetic susceptibility and
the second and fourth moments of the correlation function 
for the XY model on the square lattice
to order $\beta^{33}$ by applying the improved algorithm 
of the finite lattice method.
The long series allow us to estimate the inverse critical temperature 
as $\beta_c=1.1200(1)$, which is consistent with 
the most precise value given previously by the Monte Carlo simulation. 
The critical exponent for the multiplicative logarithmic correction 
is evaluated to be $\theta=0.054(10)$, 
which is consistent with the renormalization group prediction of $\theta=\frac{1}{16}$.
\end{abstract}

\maketitle

\section{Introduction}

It is believed that the XY model in two dimensions exhibits a phase transition 
of the  Kosterlitz-Thouless (K-T) type\cite{Kosterlitz},
which is driven by the condensation of vortices.
From the renormalization group arguments,
it was predicted that the correlation length 
has an essential singularity at the 
transition temperature $T_c$ as
\begin{equation}
   \xi \sim \exp\left(\frac{b}{t^{\sigma}}\right) ,  \label{eqn:xi_cr}
\end{equation}
with $\sigma=\frac{1}{2}$
where $t=T/T_c-1$ is the reduced temperature and $b$ is a non-universal constant.
At the critical temperature the correlation function behaves like
\begin{equation}
   G(r) \sim \frac{(\ln r)^{2\theta} }{r^{\eta}} 
      \left[ 1+O\left( \frac{\ln\ln r}{\ln r} \right) \right]  ,  
           \label{eqn:correlation_function}
\end{equation}
with $\eta=\frac{1}{4}$ and $\theta=\frac{1}{16}$,
and 
the $j$-th moment $m_{j}$ of the correlation function
behaves like
\begin{equation}
   m_{j} \sim \xi^{2+j-\eta} (\ln \xi)^{2\theta}
        \left[ 1+O\left( \frac{\ln\ln \xi}{\ln \xi} \right) \right] . \label{eqn:moment}
\end{equation}
The free energy and its temperature derivatives
(i.e., the internal energy and the specific heat)
were also predicted to behave like
\begin{equation}
   f \sim \xi^{-2} + \mbox{\rm non-singular\ term}\  .
                                                   \label{eqn:f_cr}
\end{equation}
At the critical temperature, 
the first term on the right-hand side of Eq.~(\ref{eqn:f_cr}) has 
an essential singularity with itself and its derivatives 
going to zero , while the second term stays nonzero.

The behavior in Eq.(\ref{eqn:xi_cr}) for the correlation length has been 
well established both by numerical simulations
and the high-temperature expansion.
The standard Monte Carlo simulation\cite{Edwards1991,Gupta1992} gave 
$\beta_c=1.130(15)$ with $b=2.15(10)$    
and $\beta_c=1.118(5)$ with $b=1.70(20)$  
for the square lattice. 
(Here $\beta_c$ is the critical inverse temperature, which will be defined below.)
More precise values $\beta_c=1.1208(2)$ with $b=1.800(2)$\cite{Schultka1994}  
and $\beta_c=1.1199(1)$ with $b=1.776(4)$\cite{Hasenbusch1997} were
obtained using the finite-size scaling technique 
and the renormalization group finite-size scaling method,
respectively.
In the latter approach, the renormalization group
flow of the observable was matched  
with that of the exactly solvable BCSOS model.
The latter value of $\beta_c$ was recently confirmed 
by a large scale Monte Carlo simulation
on a $2048\times 2048$ lattice using the finite-size scaling method
\cite{Hasenbusch2005}.
On the other hand the high-temperature expansion for the magnetic susceptibility 
and higher moments of the correlation function 
to order $\beta^{21}$ \cite{Campostrini1996}
and $\beta^{26}$\cite{Butera2008} 
gave less precise value $\beta_c=1.118(3)$ with $b=1.67(4)$ and 
$\beta_c=1.1198(14)$ with $b=1.77(1)$, respectively. 
It seems that the available high-temperature series is not long 
enough to give the estimation of the values for the critical temperature 
and other critical parameters in the same precision as the Monte Carlo simulation.
So it is desirable to extend the high-temperature series to much higher order.    

As for the critical exponent $\theta$ for the multiplicative logarithmic correction 
in the moments of the correlation function,
there has been controversial arguments.
Negative values ranging from $\theta=-0.077$ to 
$\theta=-0.056$ \cite{Patrascioiu1996,Janke1997,Jaster1998}
were given by the analysis of the numerical simulation 
based on the thermal scaling formula (\ref{eqn:moment}),
while the finite size scaling analysis gave positive values
in the range of 
$\theta=0.02-0.035$\cite{Kenna1997,Janke1997,Jaster1998,Chandrasekharan2003}.
The large scale Monte Carlo simulation by Hasenbusch \cite{Hasenbusch2005}
gave a value $\theta=0.056(7)$, which is consistent with the renormalization 
group prediction, 
assuming a modified finite-size-scaling behavior  
\begin{equation}
   m_{0} \sim L^{2-\eta} \left( C+ \ln L \right)^{2\theta}, \label{eqn:modified_scaling}
\end{equation}
where $L$ is the size of the lattice used in the simulation 
and $C$ is a constant. On the other hand, 
the high-temperature expansion series to order $\beta^{21}$ \cite{Campostrini1996}
gave negative values of 
$\theta=-0.042(5)$ and $\theta=-0.05(2)$ assuming the thermal scaling (\ref{eqn:moment}).    
We should note that these values from the high-temperature series 
are obtained only for the Dlog-Pad\'{e} approximants
and the general inhomogeneous differential approximants do not give 
convergent result to this order.  
Higher order series would be needed again to resolve the discrepancy 
between the result of the high-temperature series analysis and 
the renormalization group prediction.

A commonly known method for series expansions 
is the graphical method\cite{Domb1974}.
However, in this method, one must list all the graphs that 
contribute to the desired order of the series.
An alternative and powerful method to generate the expansion series is 
the finite lattice method\cite{Enting1977,Arisue1984,Creutz1991}. 
It avoids listing up all the graphs 
and it reduces the problem to the calculation of the partition functions for 
the relevant finite-size lattices, 
which is a rather straightforward procedure  
if we use the transfer matrix formulation.
In many cases, the finite lattice method generates longer series  
than the graphical method\cite{Bhanot1992,Guttmann1993,
Arisue1993,Arisue1994,Arisue1995,Bhanot1993,Guttmann1994b,
Arisue1999}.
Unfortunately, in the case of the XY model in two dimensions,
the original finite lattice method can generate 
a high-temperature series that is at most as long as the series that can be obtained 
by the graphical method.

In order to generate long series for the 
the XY model in two dimensions, we here apply an improved algorithm
of the finite lattice method developed by the author and Tabata
\cite{Arisue1995b,Arisue1995c}.
This improved algorithm is powerful in the case of models in which 
the spin variable at each site takes more than two values,
including an infinite number of values.
This algorithm was applied to generate a low-temperature series for  
the solid-on-solid model\cite{Arisue1995b} and 
high- and low-temperature series for the $q$-state Potts model 
in two dimensions\cite{Arisue1995c}. 
In both cases, it generates much longer series 
than the original finite lattice method.
The XY model in two dimensions can be mapped to a kind of 
solid-on-solid model, and the improved algorithm of the finite lattice method
in fact enabled us to obtain the high-temperature series 
for the free energy of this model on the square lattice 
to order $\beta^{48}$\cite{Arisue2007},
which is two times longer than the series previously derived.
From the analysis of the obtained long series we confirmed that
the free energy of the two-dimensional XY model behaves like Eq.~(\ref{eqn:f_cr}), 
with values of the critical temperature and the non-universal constant $b$ 
that are close to the values
obtained in the study of the correlation length.
We apply this improved algorithm of the finite lattice method 
to generate the high-temperature series to order $\beta^{33}$ 
for the magnetic susceptibility and 
the second and fourth moments of the correlation function.
The obtained long series will provide the value of the critical temperature 
in the same precision as the latest large scale Monte Carlo simulations
and the value of the critical exponent $\theta$ 
which is consistent with the renormalization group prediction.

In section 2, we describe how to apply 
the improved algorithm of the finite lattice method
to generate the high-temperature series 
for the moments of the correlation function.
In section 3, the high-temperature series to order $\beta^{33}$ are given.
Section 4 is devoted to the analysis of the obtained series
to evaluate the critical parameters.

\section{Algorithm}

 We consider the XY model defined on the square lattice. 
The Hamiltonian of this system is
\begin{equation}
      H = -\sum_{\langle i,j \rangle} J \vec{s}_i \vec{s}_j\;,
\end{equation}
where  $\vec{s}_i$ is a two-dimensional unit vector located at the lattice site $i$, 
and the summation is taken over all pairs $\langle i,j \rangle$ 
of nearest neighbor sites.
The correlation function is given by 
\begin{equation} 
      <\vec{s}_x \vec{s}_0> =  \frac{\Gamma_{x,0}}{Z} \;, \label{eqn:correlation_funcion_def}
\end{equation}
where $Z$ is the partition function
\begin{equation} \displaystyle
      Z = \int \prod_{i} d\theta_i  \exp{\left( - \frac{H}{kT} \right) }\;,
\end{equation}
and
\begin{equation} 
       \Gamma_{x,0}=
\int \prod_{i} d\theta_i  
         \vec{s}_x \vec{s}_0  \exp{\left( - \frac{H}{kT} \right) } 
                  \;. \label{eqs:correlation_funcion_def2}
\end{equation}
Here $T$ is the temperature and $\theta_i$ is the angle variable of the spin 
$\vec{s_i}=(\cos{\theta_i},\sin{\theta_i})$.

This model can be mapped exactly to a solid-on-solid model  
and the numerator and denominator of Eq.(\ref{eqs:correlation_funcion_def2})   
can be rewritten as
\begin{equation} 
Z = \sum_{\{ h\; |\, -\infty \le h_i \le +\infty  \}} 
\prod_{\langle i,j \rangle} I_{|h_i-h_j|}(\beta)\;,
\end{equation}
and 
\begin{equation} 
\Gamma_{x,0}=
       \sum_{\{ h\; |\, -\infty \le h_i \le +\infty  \}} 
\prod_{\langle i,j \rangle} I_{|h_i-h_j+\Delta_{i,j;x,0}|}(\beta)\;,
\label{eq:numerator}
\end{equation}
where $\beta=J/kT$, 
$I_n$ is the modified Bessel function,
the product is taken with respect to all the pairs of 
neighboring plaquettes,
and the variable $h_i$ at each plaquette $i$ 
takes integer value ranging from $-\infty$ to $+\infty$.
In Eq.(\ref{eq:numerator}) the $\Delta_{i,j;x,0}$ is defined as
\begin{equation} 
\Delta_{i,j;x,0}= \left\{
\begin{array}{ll} 
1 & \mbox{for}\ b_{i,j} \in L_{x,0},  \\
0 & \mbox{otherwise},  
\end{array} 
\right.
\end{equation}
where  $b_{i,j}$ is the bond sandwiched by the neighboring plaquettes $i$ and $j$,
and $L_{x,0}$ is the set of bonds on an arbitrary shortest path 
connecting the sites $0=(0,0)$ and $x=(x_1,x_2)$  along the bonds,
for which we adopt here the path
that starts from the site $0$ and go straight first in $1$ direction and 
then in $2$ direction reaching the site $x$ when $x_1 x_2\ge 0$,
and first in $2$ direction and 
then in $1$ direction when $x_1 x_2< 0$. 

 The improved algorithm of the finite lattice method 
to generate the high-temperature expansion series 
for the correlation function of this model 
is essentially the same as that for the free energy 
described in Ref.\cite{Arisue2007}.
We first calculate the correlation function 
for each of the finite-size rectangular lattice $\Lambda(l_1 \times l_2, p)$ 
with a restricted range of the value of the plaquette variable
\begin{equation}
    <\vec{s}_x \vec{s}_0>_{\Lambda ;h_{+},h_{-}}
       =\frac{\Gamma_{x,0}(\Lambda, h_{+},h_{-})}{Z(\Lambda,h_{+},h_{-})}\;.
                 \label{eqn:correlation_function_finite}
\end{equation}
Here the finite size lattice $\Lambda(l_1 \times l_2, p)$ is specified 
by its size $|\Lambda|=l_1 \times l_2$ and its position $p$, and 
each plaquette variable $h_i$ is 
restricted so that $ h_{-} \le h_i \le h_{+}$  
(where $h_{-} \le 0$ and $h_{+} \ge 0 $) both in the calculation of the 
numerator and the denominator. 
We define the size of the finite lattice so that the  
$l_1 \times l_2$ lattice involves $l_1 \times l_2$ plaquettes,
including the bonds and sites on their boundary.
For instance, the $1 \times 1$ lattice consists of a single plaquette, 
including 4 bonds and 4 sites.
We take into account finite-size lattices with $l_1=0$ and/or $l_2= 0$.
An $l_1 \times 0$ lattice consists of $l_1$ bonds and
$l_1 +1$ sites with no plaquette.
The $0 \times 0$ lattice consists only of one site with no bond or plaquette.
 The boundary condition is taken
such that all the plaquette variables outside the $l_1 \times l_2$ lattice 
are fixed to zero. 
The numerator and the denominator of Eq.(\ref{eqn:correlation_function_finite})
can be calculated efficiently by the transfer matrix method using a procedure 
in which a finite-size lattice is built 
one plaquette at a time\cite{Enting1980,Bhanot1990}.

 We then define $\phi_{x,0}(\Lambda,h_{+},h_{-})$ of the finite size lattice $\Lambda$  
and of the restricted range of the plaquette variables recursively as
\begin{eqnarray}
\!\!\!\!\!\!\!\!\!\!\!\!\!
&& \phi_{x,0}(\Lambda, h_{+},h_{-})
        = <\vec{s}_x \vec{s}_0>_{\Lambda ;h_{+},h_{-}} \nonumber\\
\!\!\!\!\!\!\!\!\!\!\!\!\!
&& 
   \quad -  \sum_{
              \genfrac{}{}{0pt}{2} 
                   {\Lambda^{\prime} \subseteq 
                    \Lambda,\ 
                    0 \le h_{+}^{\prime} \le h_{+}, \ 
                    h_{-} \le h_{-}^{\prime} \le 0 }
                   {(\Lambda^{\prime},h_{+}^{\prime},h_{-}^{\prime})
                    \ne 
                    (\Lambda, h_{+}\,h_{-})}
                  }
\!\!\!  \phi_{x,0}(\Lambda^{\prime}, h_{+}^{\prime}, h_{-}^{\prime} \}. \nonumber\\
&& 
\end{eqnarray}
It should be noted that $\phi_{x,0}(\Lambda, h_{+},h_{-})=0$ if
the site  $x$ or $0$ is not included in $\Lambda$.

The correlation function in the thermodynamic limit is then given by 
\begin{eqnarray}
   <\vec{s}_x \vec{s}_0> 
     &\equiv & \lim_{|\Lambda|\to \infty,\ h_{+}\to \infty,\ h_{-}\to -\infty} 
                <\vec{s}_x \vec{s}_0>_{\Lambda ;h_{+},h_{-}} \nonumber\\
     &=&      \sum_{\Lambda,h_{+},h_{-}} 
                  \phi_{x,0}(\Lambda,h_{+},h_{-}) . 
      \label{eqn:correlation_function_infty}
\end{eqnarray}
In the last line of Eq.(\ref{eqn:correlation_function_infty}) the summation
should be taken for all the lattice size and all of its position and 
all integer values of $h_{+}$ and $h_{-}$ 
with $0\le h_{+}\le \infty$, $-\infty\le h_{-}\le 0$.

\begin{figure}[tb!]
\includegraphics[scale=0.75]{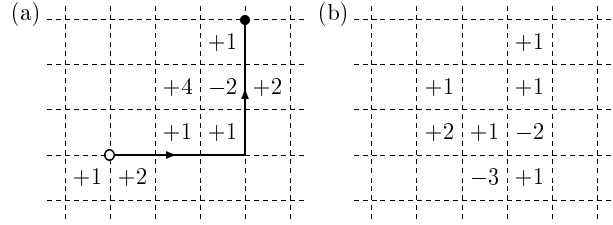}
\caption{
 (a) Example of the main polymer in the standard cluster expansion.
 The open circle indicates the site $(0,0)$ 
 where the spin $\vec{s}_0$ exists
 and the closed circle indicates the site $x=(x_1,x_2)$ 
 where  the spin $\vec{s}_x$ exists.
 (b) Example of the sub-polymer. 
}
\label{figure:polymer1}
\end{figure}

 In the standard (graphical) cluster expansion of the correlation function 
for the SOS model, 
a cluster is composed of polymers: one main polymer that 
consists of the set of bonds $L_{x,0}$ and connected plaquettes  
attaching to it and possible sub-polymers that consists of connected plaquettes. 
An example of the main polymer and sub-polymer can be seen in Fig.~1.
A value $h_i\ ( \neq 0 )$ is assigned to each plaquette $i$ of the polymer. 
 We can assign to each cluster 
two numbers, $h_{\rm max}\ (\ge 0)$ and $h_{\rm min}\ (\le 0)$, 
which are the maximum and the minimum, respectively, 
of the plaquette variable $h_i$
in all the plaquettes of the polymers of which the cluster consists.
 Then, we can prove\cite{Arisue1984} 
that $\phi_{x,0}(\Lambda, h_{+},h_{-})$ includes 
the contributions to $<\vec{s}_x \vec{s}_0>$ from all the clusters of polymers 
in the standard cluster expansion 
for which $h_{\rm max}=h_{+}$ and $h_{\rm min}=h_{-}$ 
and that can be embedded into the lattice $\Lambda$ 
but cannot be embedded into any of its rectangular sub-lattices.

\begin{figure}[tb!]
\includegraphics[scale=0.75]{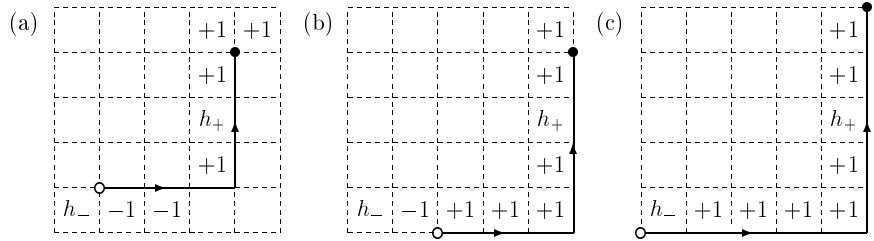}
\caption{
Examples of the cluster composed of a single main polymer that contributes to the 
lowest order term of the high-temperature expansion of $\phi(\Lambda,h_{+},h_{-})$ 
for $h_{+}\ge 1$ and $h_{-}\le -1$. 
The size of the finite lattices in these examples is $|\Lambda|=5\times 5$.  
}
\label{figure:cluster1}
\end{figure}

\begin{figure}[tb!]
\includegraphics[scale=0.8]{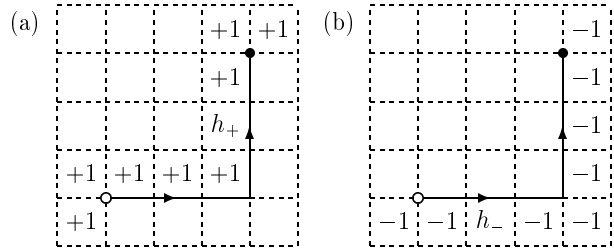}
\caption{
Examples of the cluster composed of a single main polymer that contributes to the 
lowest order term of the high-temperature expansion of $\phi(\Lambda,h_{+},h_{-})$
(a) for $h_{+}\ge 1$ and $h_{-}=0$ and (b) for $h_{+}=0$ and $h_{-}\le -1$.
The size of the finite lattice in these examples is $|\Lambda|=5\times 5$.  
}
\label{figure:cluster2}
\end{figure}

Now we consider from what order the series expansion  
of $\phi_{x,0}(\Lambda,h_{+},h_{-})$ with respect to $\beta$ starts.  
It is enough to give the order for the position $x=(x_1,x_2)$ 
so that $x_1\ge 0$ and $x_2\ge 0$.
The order for the other cases is known 
by the $90^{\circ}$ rotational symmetry of the model.
Any cluster that contributes to the lowest-order term 
of the series expansion 
for $\phi_{x,0}(\Lambda,h_{+},h_{-})$ consists only of a main polymer.
Hence the series expansion of $\phi_{x,0}(\Lambda,h_{+},h_{-})$  
begins 
from order $\beta^{\, n_{x,0}(\Lambda,h_{+},h_{-})}$ 
in the case of $h_{+} \ge 1$ and $h_{-} \le -1$ with 
\begin{equation} \! \! \! \! \! \! \! \! \! \! \! \!  
  n_{x,0}(\Lambda, h_{+},h_{-}) = 
\left\{
\begin{array}{l}
   2l_1 + 2l_2 -x_1-x_2 + 4h_{+}+4|h_{-}|-4 \\ \qquad  
      \mbox{for}\ (p_1,p_2)=(0,0)\ 
      \mbox{and}\ (p^{\prime}_1,p^{\prime}_2)=(x_1,x_2), \\
   2l_1 + 2l_2 -x_1-x_2 + 4h_{+}+4|h_{-}|-6 \\ \qquad 
      \mbox{for}\ p_2=0,\ p^{\prime}_1=x_1, \   
      \mbox{and}\ (p_1,p^{\prime}_2)\ne (0,x_2) \\
   2l_1 + 2l_2 -x_1-x_2 + 4h_{+}+4|h_{-}|-8  \\ \qquad  
      \mbox{for the others}. 
                                           \label{eqn:order_to_start1}
\end{array}
\right.
\end{equation}
Here we denote $(p_1,p_2)$ as the position of the bottom-left corner 
and $(p^{\prime}_1,p^{\prime}_2)$ as the position of the top-right corner, respectively,
of the lattice $\Lambda$ with its size $l_1\times l_2$
($p^{\prime}_1-p_1=l_1$ and $p^{\prime}_2-p_2=l_2$). 
Examples of the main polymer are given in Fig.~2 (a) (b) (c),
which correspond to the three cases in Eq.(\ref{eqn:order_to_start1}), respectively.
In the case of $h_{+} \ge 1$ and $h_{-}= 0$ we have
\begin{equation} \! \! \! \! \! \! \! \! \! \! \! \!\! \! \!  
  n_{x,0}(\Lambda, h_{+},h_{-}) = 
\left\{
\begin{array}{l}
   2l_1 + 2l_2 -x_1-x_2 + 4h_{+} \\ \qquad  
      \mbox{for}\ p_1=0,\ p^{\prime}_2=x_2 \ 
      \mbox{and}\ (p_2,p^{\prime}_1)\ne (0,x_1), \\
   2l_1 + 2l_2 -x_1-x_2 + 4h_{+} -2 \\ \qquad  
      \mbox{for}\ p_1=0,\ p_2<0 \ 
      \mbox{and}\ (\ p^{\prime}_2>x_2\ \mbox{or} \ p^{\prime}_1=x_1), \\
   2l_1 + 2l_2 -x_1-x_2 + 4h_{+} -2 \\ \qquad  
      \mbox{for}\ p^{\prime}_2=x_2,\ p^{\prime}_1>x_1\\ \qquad \ \ \ \ 
      \mbox{and}\ (\ p_1<0\ \mbox{or}\ \ (p_1,p_2)=(0,0)), \\
   2l_1 + 2l_2 -x_1-x_2 + 4h_{+} -4  \\ \qquad  
      \mbox{for the others},
                                           \label{eqn:order_to_start2}
\end{array}
\right.
\end{equation}
and in the case of $h_{+} = 0$ and $h_{-}\le  -1$ 
\begin{equation} \! \! \! \! \! \! \! \! \! \! \! \!\! \! \!  
  n_{x,0}(\Lambda, h_{+},h_{-}) = 
\left\{
\begin{array}{l}
   2l_1 + 2l_2 -x_1-x_2 +4|h_{-}|\\ \qquad  
      \mbox{for}\ (p_1,p_2)=(0,0)\ 
      \mbox{and}\ (p^{\prime}_1,p^{\prime}_2)=(x_1,x_2), \\
   2l_1 + 2l_2 -x_1-x_2 +4|h_{-}|+4\\ \qquad  
      \mbox{for}\ p_2=0,\ p^{\prime}_1=x_1 \ 
      \mbox{and}\ (p_1,p^{\prime}_2)\ne (0,x_2), \\
   2l_1 + 2l_2 -x_1-x_2 +4|h_{-}| \\ \qquad
      \mbox{for}\ p_2=0,\ p_1<0 \  
      \mbox{or}\ p^{\prime}_1=x_1,\ p^{\prime}_2>x_2,\\
   2l_1 + 2l_2 -x_1-x_2 +4|h_{-}|-4  \\ \qquad  
      \mbox{for the others}. 
                                           \label{eqn:order_to_start3}
\end{array}
\right.
\end{equation}
Examples of the main polymer are given in Fig.~3 (a) and (b),
which correspond to the last case in Eq.(\ref{eqn:order_to_start2})
and the last case in Eq.(\ref{eqn:order_to_start3}), respectively.
In the case of $h_{+} = 0$ and $h_{-}=0$ we have
\begin{equation}   
  n_{x,0}(\Lambda, h_{+},h_{-}) = 
\left\{
\begin{array}{l}
   l_1 + l_2  \ \   \mbox{for}\ (p_1,p_2)=(0,0)\ 
   \mbox{and}\ (p^{\prime}_1,p^{\prime}_2)=(x_1,x_2), \\
   \infty  \ \ \ \ \ \ \ \mbox{for the others} .
                                           \label{eqn:order_to_start4}
\end{array}
\right.
\end{equation}
Here $n_{x,0}(\Lambda, h_{+},h_{-}) = \infty$ implies 
that $\phi_{x,0}(\Lambda, h_{+},h_{-})=0 $. 

Thus, in order to obtain the expansion series 
for the correlation function $<\vec{s}_x \vec{s}_0>$ to order $\beta^N$,
we have only to take into account all combinations of the rectangular lattice
$\Lambda$ and the range of the plaquette variable $(h_{+},h_{-})$ 
that satisfy the relation 
$ n_{x,0}(\Lambda,h_{+},h_{-}) \le N $
in the summation of Eq.~(\ref{eqn:correlation_function_infty}) 
and to evaluate each of the $\phi_{x,0}(\Lambda,h_{+},h_{-})$ to order $\beta^N$.

The $j$-th moment of the correlation function is given by
\begin{equation}
  m_{j}=\sum_{x} |x|^{j} <\vec{s}_x \vec{s}_0> \label{eqn:m_j}
\end{equation}
where $|x|$ is the distance between the site $x$ and $0$.
The moment can be calculated more efficiently in the following way.
First we calculate
$\tilde{\Gamma}_{j}(l_1,l_2,h_{+},h_{-})$ defined by
\begin{equation}
 \tilde{\Gamma}_{j}(l_1,l_2,h_{+},h_{-}) 
     =    \frac{\displaystyle \sum_{x,y\subseteq \Lambda}
            |x-y|^{j} \, \Gamma_{x,y} \left(\Lambda,h_{+},h_{-}\right)}
              {Z\left( \Lambda \right)} \label{eq:Gamma_tilde}
\end{equation}
for each finite size lattice $\Lambda$.
We note that the numerator and the denominator of Eq.(\ref{eq:Gamma_tilde}) 
depend only on the lattice size 
$|\Lambda|=l_1\times l_2$ and on $h_{+}, h_{-}$ 
and that they are independent of the position of the lattice 
after the summation is taken for $x$ and $y$. 
They can be calculated efficiently again
by the transfer matrix method using the procedure in which 
a finite-size lattice is built one plaquette at a time
not only for $j=0$ but also for $j\ge2$ ($j$ : even), the detail of which 
was described in the application of the finite lattice method 
to the calculation of the low-temperature series 
for the second moment of the correlation function 
in the simple cubic Ising model \cite{Arisue1995}.
The moment is then given by  
\begin{equation}
  m_{j}=\sum_{l_1,l_2,h_{+},h_{-}} 
                  \tilde{\phi}_{j}(l_1,l_2,h_{+},h_{-})
\end{equation}
where $\tilde{\phi}_{j}(l_1,l_2,h_{+},h_{-})$ is defined recursively as
\begin{eqnarray}
\!\!\!\!\!\!\!\!\!\!\!\!\!\!\!\!\!\!
&&  \tilde{\phi}_{j}(l_1,l_2,h_{+},h_{-})
        =\tilde{\Gamma}_{j}(l_1,l_2,h_{+},h_{-}) \nonumber \\
\!\!\!\!\!\!\!\!\!\!\!\!\!\!\!\!\!\!
&&  - \sum_{
         \genfrac{}{}{0pt}{2} 
          { l_1^{\prime},l_2^{\prime},h_{+}^{\prime},h_{-}^{\prime} }
          { (l_1^{\prime},l_2^{\prime},h_{+}^{\prime},h_{-}^{\prime})\ne  
            (l_1,l_2,h_{+},h_{-})  }
            }
       (l_1-l_1^{\prime}+1)(l_2-l_2^{\prime}+1)
       \tilde{\phi}_{j}(l_1^{\prime},l_2^{\prime},h_{+}^{\prime},h_{-}^{\prime}) . 
                       \label{eqn:phi_tilde} 
\end{eqnarray}
From Eq.(\ref{eqn:order_to_start1}) - (\ref{eqn:order_to_start4}) we know that
$\tilde{\phi}_{j}(l_1,l_2,h_{+},h_{-})$ starts from order $\beta^{\tilde{n}(l_1,l_2,h_{+},h_{-})}$
with
\begin{equation}
  \tilde{n}(l_1,l_2,h_{+},h_{-})=\left\{
\begin{array}{l}
         l_1+l_2
             \ \ \  \mbox{for}\  h_{+}=h_{-}=0,  \\ 
         l_1+l_2+4h_{+}-4
             \ \ \  \mbox{for}\  h_{+}>0 \ \mbox{and}\ h_{-}=0, \\
         l_1+l_2+4|h_{-}|-3
             \ \ \  \mbox{for}\  h_{+}=0 \ \mbox{and}\ h_{-}<0,  \\
         l_1+l_2+4h_{+}+4|h_{-}|-7
             \ \ \  \mbox{for}\  h_{+}>0 \ \mbox{and}\  h_{-}<0.  
\end{array}
\right.
          \label{eqn:order_to_start_tilde}
\end{equation}


\begin{table}[tb!]
\caption{Coefficients of the high-temperature series for the magnetic susceptibility $m_0$.}
\label{table:coefficients_chi}
{\scriptsize 
\begin{center}
\begin{tabular}{r|rcl}
    $n$  & &   $a^{(0)}_n$  & \\
\hline
$   0 $ &\ $                                  1 $ & / &  $                       1 $ \\
$   1 $ &\ $                                  4 $ & / &  $                       1 $ \\
$   2 $ &\ $                                 12 $ & / &  $                       1 $ \\
$   3 $ &\ $                                 34 $ & / &  $                       1 $ \\
$   4 $ &\ $                                 88 $ & / &  $                       1 $ \\
$   5 $ &\ $                                658 $ & / &  $                       3 $ \\
$   6 $ &\ $                                529 $ & / &  $                       1 $ \\
$   7 $ &\ $                              14933 $ & / &  $                      12 $ \\
$   8 $ &\ $                               5737 $ & / &  $                       2 $ \\
$   9 $ &\ $                             389393 $ & / &  $                      60 $ \\
$  10 $ &\ $                            2608499 $ & / &  $                     180 $ \\
$  11 $ &\ $                            3834323 $ & / &  $                     120 $ \\
$  12 $ &\ $                            1254799 $ & / &  $                      18 $ \\
$  13 $ &\ $                           84375807 $ & / &  $                     560 $ \\
$  14 $ &\ $                         6511729891 $ & / &  $                   20160 $ \\
$  15 $ &\ $                        66498259799 $ & / &  $                   96768 $ \\
$  16 $ &\ $                      1054178743699 $ & / &  $                  725760 $ \\
$  17 $ &\ $                     39863505993331 $ & / &  $                13063680 $ \\
$  18 $ &\ $                     19830277603399 $ & / &  $                 3110400 $ \\
$  19 $ &\ $                   8656980509809027 $ & / &  $               653184000 $ \\
$  20 $ &\ $                   2985467351081077 $ & / &  $               108864000 $ \\
$  21 $ &\ $                 811927408684296587 $ & / &  $             14370048000 $ \\
$  22 $ &\ $                 399888050180302157 $ & / &  $              3448811520 $ \\
$  23 $ &\ $              245277792666205990697 $ & / &  $           1034643456000 $ \\
$  24 $ &\ $               83292382577873288741 $ & / &  $            172440576000 $ \\
$  25 $ &\ $              376988970189597090587 $ & / &  $            384296140800 $ \\
$  26 $ &\ $            62337378385915430773643 $ & / &  $          31384184832000 $ \\
$  27 $ &\ $           480555032864478422139959 $ & / &  $         119830523904000 $ \\
$  28 $ &\ $            80636088313579215330647 $ & / &  $           9985876992000 $ \\
$  29 $ &\ $      19240186846097940775812460721 $ & / &  $     1186322186649600000 $ \\
$  30 $ &\ $      34266867760374182809422566317 $ & / &  $     1054508610355200000 $ \\
$  31 $ &\ $    9864232002328615891762221069959 $ & / &  $   151849239891148800000 $ \\
$  32 $ &\ $      93697376428024822547085555709 $ & / &  $      723091618529280000 $ \\
$  33 $ &\ $  665861878626519700317398249291449 $ & / &  $  2581437078149529600000 $ \\
\hline
\end{tabular}             
\end{center}
}
\end{table}
\begin{table}[tb!]
\caption{Coefficients of the high-temperature series 
for the second moment $m_2$ of the correlation function.}
\label{table:coefficients_2ndmom }
{\scriptsize 
\begin{center}
\begin{tabular}{r|rcl}
    $n$  & &   $a^{(2)}_n$  & \\
\hline
$   0 $ & $                                     0 $  & / & $                       1  $ \\
$   1 $ & $                                     4 $  & / & $                       1  $ \\
$   2 $ & $                                    32 $  & / & $                       1  $ \\
$   3 $ & $                                   162 $  & / & $                       1  $ \\
$   4 $ & $                                   672 $  & / & $                       1  $ \\
$   5 $ & $                                  7378 $  & / & $                       3  $ \\
$   6 $ & $                                 24772 $  & / & $                       3  $ \\
$   7 $ & $                                312149 $  & / & $                      12  $ \\
$   8 $ & $                                 77996 $  & / & $                       1  $ \\
$   9 $ & $                              13484753 $  & / & $                      60  $ \\
$  10 $ & $                              28201211 $  & / & $                      45  $ \\
$  11 $ & $                             611969977 $  & / & $                     360  $ \\
$  12 $ & $                             202640986 $  & / & $                      45  $ \\
$  13 $ & $                           58900571047 $  & / & $                    5040  $ \\
$  14 $ & $                            3336209179 $  & / & $                     112  $ \\
$  15 $ & $                         1721567587879 $  & / & $                   23040  $ \\
$  16 $ & $                        16763079262169 $  & / & $                   90720  $ \\
$  17 $ & $                      5893118865913171 $  & / & $                13063680  $ \\
$  18 $ & $                     17775777329026559 $  & / & $                16329600  $ \\
$  19 $ & $                   1697692411053976387 $  & / & $               653184000  $ \\
$  20 $ & $                     41816028466101527 $  & / & $                 6804000  $ \\
$  21 $ & $                 206973837048951639371 $  & / & $             14370048000  $ \\
$  22 $ & $                 721617681295019782781 $  & / & $             21555072000  $ \\
$  23 $ & $               79897272060888843617033 $  & / & $           1034643456000  $ \\
$  24 $ & $                2287397511857949924319 $  & / & $             12933043200  $ \\
$  25 $ & $             5412508223507386985733313 $  & / & $          13450364928000  $ \\
$  26 $ & $             7139182711315236460182251 $  & / & $           7846046208000  $ \\
$  27 $ & $           107851995064346070336358789 $  & / & $          52725430517760  $ \\
$  28 $ & $            75355895214528595226953733 $  & / & $          16476697036800  $ \\
$  29 $ & $       1724076192091313972941261252343 $  & / & $      169474598092800000  $ \\
$  30 $ & $      53429382179619216000735913825117 $  & / & $     2372644373299200000  $ \\
$  31 $ & $    7534609247991680570113055733178247 $  & / & $   151849239891148800000  $ \\
$  32 $ & $     344373363823985360326961701207501 $  & / & $     3163525831065600000  $ \\
$  33 $ & $   68217033997582723452684940622596049 $  & / & $   286826342016614400000  $ \\
\hline
\end{tabular}             
\end{center}
}
\end{table}
\begin{table}[tb!]
\caption{Coefficients of the high-temperature series 
for the fourth moment $m_4$ of the correlation function.}
\label{table:coefficients_4thmom}
{\scriptsize 
\begin{center}
\begin{tabular}{r|rcl}
    $n$  &    &   $a^{(4)}_n$  &  \\
\hline
$   0 $ & $                                          0  $  & / & $                        1  $ \\
$   1 $ & $                                          4  $  & / & $                        1  $ \\
$   2 $ & $                                         96  $  & / & $                        1  $ \\
$   3 $ & $                                        930  $  & / & $                        1  $ \\
$   4 $ & $                                       6112  $  & / & $                        1  $ \\
$   5 $ & $                                      96850  $  & / & $                        3  $ \\
$   6 $ & $                                     147648  $  & / & $                        1  $ \\
$   7 $ & $                                    7305173  $  & / & $                       12  $ \\
$   8 $ & $                                    2319540  $  & / & $                        1  $ \\
$   9 $ & $                                  498173873  $  & / & $                       60  $ \\
$  10 $ & $                                 1271029508  $  & / & $                       45  $ \\
$  11 $ & $                                 2210163319  $  & / & $                       24  $ \\
$  12 $ & $                                 2606525954  $  & / & $                        9  $ \\
$  13 $ & $                              4449953438647  $  & / & $                     5040  $ \\
$  14 $ & $                              3300804041221  $  & / & $                     1260  $ \\
$  15 $ & $                             81597010130527  $  & / & $                    10752  $ \\
$  16 $ & $                           1952458419167723  $  & / & $                    90720  $ \\
$  17 $ & $                         782261299458533011  $  & / & $                 13063680  $ \\
$  18 $ & $                         222847854860540393  $  & / & $                  1360800  $ \\
$  19 $ & $                      287995465331747559427  $  & / & $                653184000  $ \\
$  20 $ & $                       15925021359550756357  $  & / & $                 13608000  $ \\
$  21 $ & $                     8810941751514830879983  $  & / & $               2874009600  $ \\
$  22 $ & $                     8551911058786623741001  $  & / & $               1077753600  $ \\
$  23 $ & $                 21013548184238208070598633  $  & / & $            1034643456000  $ \\
$  24 $ & $                  1108989303528609120610577  $  & / & $              21555072000  $ \\
$  25 $ & $                 31566975982648926005193559  $  & / & $             244552089600  $ \\
$  26 $ & $                629546809875204592961945533  $  & / & $            1961511552000  $ \\
$  27 $ & $            1043070737244084798242371813021  $  & / & $         1318135762944000  $ \\
$  28 $ & $              10631253805062761391056806729  $  & / & $            5492232345600  $ \\
$  29 $ & $         5575566587216452758709818963741041  $  & / & $      1186322186649600000  $ \\
$  30 $ & $         2240485100208335200628893992637883  $  & / & $       197720364441600000  $ \\
$  31 $ & $        63403925202711962924615730240227527  $  & / & $      2336142152171520000  $ \\
$  32 $ & $         8174296006615548272296498779225887  $  & / & $       126541033242624000  $ \\
$  33 $ & $    394547895932847211470169206205154958361  $  & / & $   2581437078149529600000  $ \\
\hline
\end{tabular}             
\end{center}
}
\end{table}


\section{Series}
 We have calculated the high-temperature expansion series to order $\beta^{33}$ 
for the $j$-th moments $m_{j}\ (j=0,2,4)$ of the
correlation function for the XY model on the square lattice 
($m_0$ is the magnetic susceptibility).
 The obtained expansion coefficients are listed 
in Table~I, II and III, 
where the coefficient $a^{(j)}_n$ is defined as
\begin{equation}
  m_{j} = \sum_{n=1}^{N} a^{(j)}_n \left(\frac{\beta}{2}\right)^n.  \label{eqfe}
\end{equation}
 We have checked that each of the $\tilde{\phi}_{j}(l_1, l_2;h_{+},h_{-})$'s
in Eq.~(\ref{eqn:phi_tilde}) starts from the correct order in $\beta$, 
as given by Eq.~(\ref{eqn:order_to_start_tilde}).
 Our series coincide exactly
with the series to order $\beta^{21}$ for $j=0,2$ and $4$ 
given by Campostrini et al.\cite{Campostrini1996}
and to order $\beta^{26}$ for $j=0$ and $2$ 
given by Butera and Pernici\cite{Butera2007,Butera2008},
which was obtained by a graphical method
and a non-graphical recursive algorithm based on the Schwinger-Dyson equations,
respectively. 

To obtain the large numerators and denominators of the coefficients exactly,
we have used the same technique as  
in the calculation of the free energy series\cite{Arisue2007}.
In each step of the calculation the series expansion of a function $f(\beta)$ is
expressed as
\begin{equation}
   f(\beta)= \sum_n \frac{a_{n}}{n!} \left(\frac{\beta}{2}\right)^n,\ \ \ 
    g(\beta)= \sum_n \frac{b_{n}}{n!} \left(\frac{\beta}{2}\right)^n,
                                         \label{eqn:Bessel}
\end{equation}
then the product of the two functions is given by
\begin{equation}
   f(\beta )g(\beta ) 
    = \sum_n \frac{c_n}{n!} \left(\frac{\beta}{2}\right)^n\;,
                                         \label{eqn:product}
\end{equation}
with
\begin{equation}
   {c}_n= \sum_{n'=0}^{n} \frac{n!}{n'!(n-n')!} a_{n'} b_{n-n'}\;,
\end{equation}
and, if $a_n$'s and $b_n$'s are integers, $c_n$'s are also integers.

The calculations were carried out on a PC cluster
at the Information Processing Center at OPCT and
on an Altix3700 BX2 at Yukawa Institute of Kyoto University.

\section{Series Analysis}
From Eq.(\ref{eqn:moment}) the logarithm of the moment of the correlation function 
is expected to behave near the critical temperature like
\begin{equation}
  \ln m_{j} \sim \frac{(2+j-\eta ) b}{\tau^{\sigma}} + O(\ln \tau) , \label{eqn:log_m_j}
\end{equation}
where $\tau =1-\beta/\beta_c$ and $\beta_c=J/kT_c$ is the critical inverse temperature.
Here we analyze it by the first order 
inhomogeneous differential approximation(IDA), in which the differential equation 
for $f(\beta)=\ln m_{j}$ is satisfied  as
\begin{equation}
  Q_m(\beta ) f^{\prime}(\beta ) + P_l(\beta ) f(\beta ) 
                                +R_k(\beta )=O( \beta^{m+l+k+2} ),
\end{equation}
where $Q_m(\beta)$, $P_l(\beta)$ and $R_k(\beta)$ are polynomials of order $m$, $l$ and $k$ 
respectively and $Q_m(0)=1$.
The critical inverse temperature $\beta_c$ is given by the zero of $Q_m(\beta)$
and the exponent $\sigma$ is evaluated by 
\begin{equation}
  \sigma=-\frac{P_l(\beta_c)}{Q'_m(\beta_c)}.
\end{equation}
In the analysis by the first order IDA here and below,
we restrict $m+l+k+2$ to be the maximum order of the analyzed series 
with $-1\le k \le 9$ and $|m-l|\le 4$, and exclude
the approximants that have another zero of $Q_m(\beta)$  
with $|\beta-\beta_c|/\beta_c<0.10$,
which is called near-by singularity.

\begin{figure}[tb!]
\includegraphics[scale=0.68]{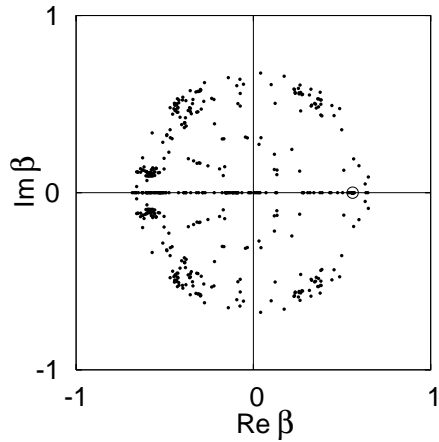}
\caption{
Plot of poles in the unbiased inhomogeneous deferential approximation for $\ln m_0$.
The open circle indicates the poles corresponding to the physical singularity .
}
\label{figure:poles}
\end{figure}

In Fig.~4 we plot the zeros of $Q_m(\beta)$ in the complex plane of $\beta$.
Besides clear accumulation of the points around $\beta=1.12$ we see
another accumulation around the nonphysical point $\beta=\beta_0\sim -1.2\pm 0.1 i$.
In order to remove the influence of this nonphysical critical point,
we have made Euler transformation 
\begin{equation}
 \beta'=\frac{\beta}{1-\beta/\beta_0}.
\end{equation}
with $\beta_0=-1.2$. 
In fact after this transformation 
the IDA gives a series of ($\beta_c$, $\sigma$) 
which converges better onto a straight line. 
Biased analysis fixing $\sigma=\frac{1}{2}$ gives $\beta_c=1.1200(4)$.
Hence we apply this Euler transformation 
in all of the series analysis presented below.  

\begin{figure}[tb!]
\includegraphics[scale=0.68]{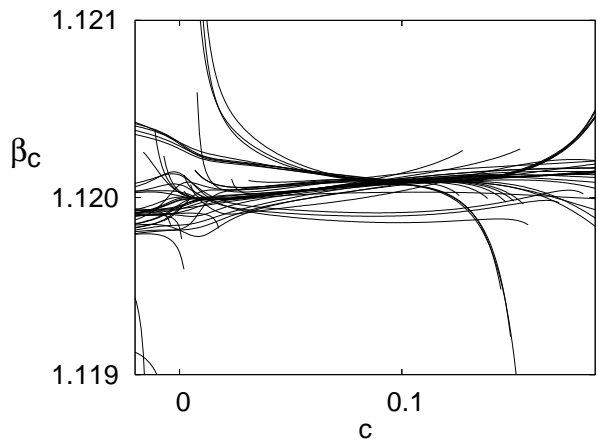}
\caption{
$\beta_c$ given by the biased IDA 
for $\ln m_0 + c \ln \frac{m_2}{4\beta}$.
}
\label{figure:beta_c_biased_sigma}
\end{figure}

The above analysis ignores the existence of the sub-leading correction terms
to the leading power-law singularity of $\ln m_{j}$ in Eq.(\ref{eqn:log_m_j}).
So we have analyzed a combination of $m_0$ and $m_2$ as
\begin{equation}
  \ln m_0 + c\, \ln \frac{m_2}{4\beta}  \label{eqn:lenear_comb},
\end{equation}
searching the parameter $c$ which will give more convergent result,
anticipating that the sub-leading terms 
will be cancelled between the first and
second terms in Eq.(\ref{eqn:lenear_comb}). 
In Fig.~5 we plot the estimated value of $\beta_c$ versus $c$ in the biased analysis fixing $\sigma=\frac{1}{2}$.
We find that the choice of the parameter $c=0.08-0.12$ gives nicely converging result
of $\beta_c=1.12007(4)$.
We have also analyzed another combination as 
\begin{equation}
  \ln m_0 + c'\, \ln \left(1+ c''\, m_2 \right),
\end{equation}
where $c^{\prime\prime}$ is an arbitrary parameter which should be chosen 
so that the combination will give the most convergent result.
We have obtained the most convergent result of 
$\beta_c=1.12003(5)$ for $c'=0.05$ and $c''=0.58$
and $\beta_c=1.11997(5)$ for $c'=0.03$ and $c''=0.68$.
Thus we can safely estimate $\beta_c=1.1200(1)$.
This value is quite consistent with 
the most precise value $\beta_c=1.1199(1)$ obtained by Hasenbusch
from the large-scale Monte Carlo simulation\cite{Hasenbusch2005}.
Our value is much more precise than the value
$\beta_c=1.1198(14)$ by Butera and Pernici\cite{Butera2007,Butera2008}
from the high-temperature series to order $\beta^{26}$ .

Assuming the critical behavior of Eq.(\ref{eqn:xi_cr}) and (\ref{eqn:moment})
we can estimate the non-universal parameter $b$ by the Pad\'{e} approximation of 
\begin{equation}
\left(1-\frac{\beta}{\beta_c}\right)^{\frac{1}{2}} 
\ln\left(1+c\, \frac{m_2}{m_0}\right) \sim b
\end{equation}
and 
\begin{equation}
\left(1-\frac{\beta}{\beta_c}\right)^{\frac{1}{2}} \ln\left(1+c^{\prime}\, \frac{m_4}{m_2}\right) \sim b. 
\end{equation}
By searching the parameter $c$ or $c'$ that gives the most convergent estimation of $b$
keeping $\beta_c=1.12000$ and $\displaystyle \sigma=\frac{1}{2}$,
both of the two give the same result of $b=1.758(1)$
for $c=3.15-3.23$ and for $c^{\prime}=0.806-0.812$ respectively.
Here we have used all of $[m,l]$ Pad\'{e} approximants with $m\ge 14$ and $l\ge 14$. 
This value is a bit smaller than 
$b=1.800(2)$\cite{Schultka1994} and $b=1.776(4)$\cite{Hasenbusch1997} 
obtained in the Monte Carlo simulations. 

Unfortunately the long series do not improve the estimation 
of the exponent $\eta$ so much.
For instance, the Pad\'{e} approximation of
the quantities  
\begin{equation}
\frac{\ln\left(1+c{\displaystyle \frac{m_2}{m_0}}\right)}{\ln{m_0}} 
  \sim \frac{2}{2-\eta}
\end{equation}
and
\begin{equation}
\frac{\ln\left(1+c^{\prime}{\displaystyle \frac{m_2}{m_0^2}}\right)}{\ln{m_0}} 
  \sim \frac{\eta}{2-\eta}
\end{equation}
give the most convergent result of 
$\eta=0.256(2)$ for $c=0.69$ and $\eta=0.227(2)$ for $c^{\prime}=0.60$, respectively. 
The Pad\'{e} approximation of
\begin{equation}
\left(1-\frac{\beta}{\beta_c}\right)^{\frac{1}{2}}
\ln\left(1+c^{\prime\prime}\, \frac{m_2}{m_0^2}\right) 
                \sim  \eta b
\end{equation}
gives $\eta b=0.429(5)$ for $c^{\prime\prime}=0.83-0.89$ and, 
if we combine this with the above result $b=1.758(1)$, 
gives $\eta=0.244(3)$

\begin{figure}[tb!]
\includegraphics[scale=0.68]{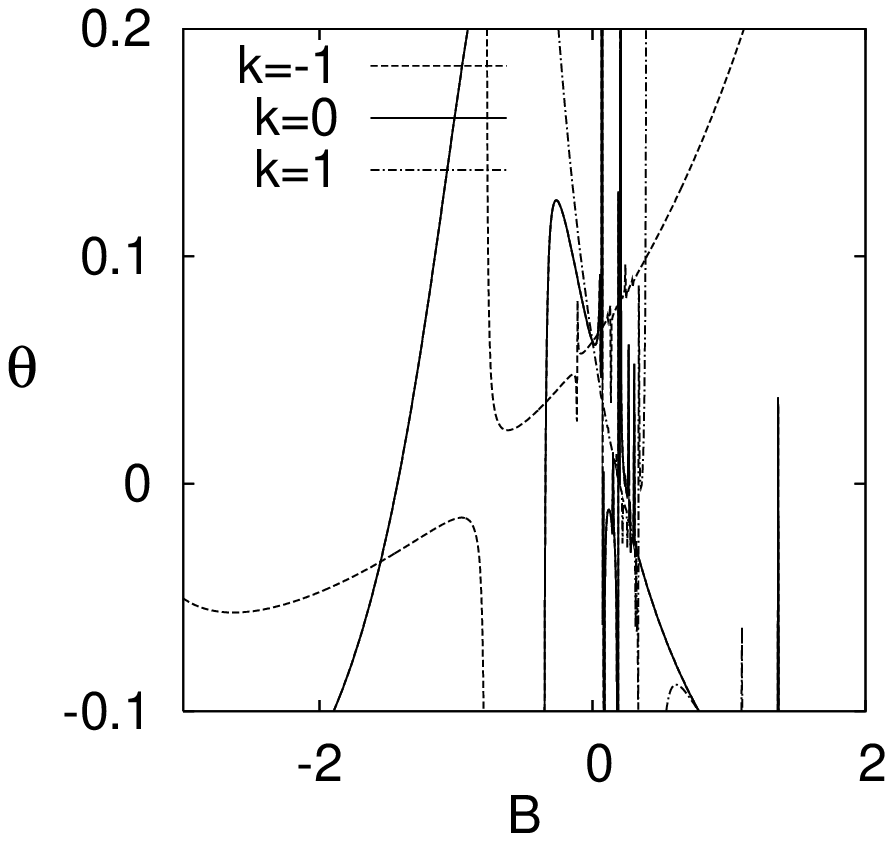}
\caption{
Plot of the estimated value of $\theta$ 
by the 1st order IDA for the test function versus $B$.
}
\label{figure:test_ida_1}
\end{figure}

As for the exponent $\theta$, Dlog-Pad\'{e} analysis of the 20th order 
series of the quantities
\begin{equation}
  \frac{m_0}{{\left( \displaystyle \frac{m_2}{m_0} \right)}^{2-\eta}} 
  \sim \frac{m_4}{{\left( \displaystyle \frac{m_2}{m_0} \right)}^{6-\eta}} 
  \sim \tau^{-2\sigma \theta}\{1+O(\tau^{\sigma} \ln \tau)\} \label{eqn:non-combined}
\end{equation}
gave estimations $\theta=-0.042(5)$ and $\theta=-0.05(2)$\cite{Campostrini1996},
which have the sign opposite to the renormalization group prediction of $\theta=\frac{1}{16}$. 
The situation does not change even if our long series are used 
in the Dlog-Pad\'{e} analysis of these quantities. 
The series of the two quantities to 33rd order 
gives $\theta=-0.019(1)$ and $\theta=-0.015(9)$, respectively.
The IDA with $k\ge 0$ gives rather convergent values within the same $k$
but quite scattered values for different $k$'s.

One possible reason why IDA for these quantities give scattered
values for different $k$ 
may be the existence of the sub-leading 
logarithmically singular term in Eq.(\ref{eqn:non-combined}),
which comes from the correction factor in Eq.(\ref{eqn:moment}).
In fact the sub-leading logarithmic singularity can strongly disturb 
the correct evaluation of the leading power low exponent $2\sigma\theta$ 
if this exponent is as small as $\frac{1}{16}$. 
We plot in Fig.~6 the estimated value of $\theta$ 
by IDA for the expansion series to order $\beta^{33}$ for a test function
\begin{equation}
    \tau^{-2\sigma\theta} \left\{1+  \tau^{\sigma}(A + B \ln \tau)\right\} \label{eqn:test_function}
\end{equation}
with $\theta=\frac{1}{16}$, $\sigma=\frac{1}{2}$, and $A=0$ plotted versus $B$.
We find that the estimated value of $\theta$ is quite sensitive to the amplitude $B$ 
of the logarithmic term, and although each approximant with the same $k$ gives
rather convergent result for any fixed value of $B$,
the approximants with different $k$ give the estimation of $\theta$
far from each other. 

\begin{figure}[tb!]
\includegraphics[scale=0.68]{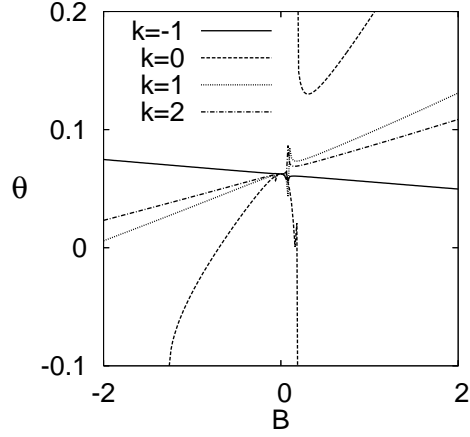}
\caption{
Plot of the estimated value of $\theta$ 
by the 2nd order IDA for the test function versus $B$.
}
\label{figure:test_ida_2}
\end{figure}

\begin{figure}[tb!]
\includegraphics[scale=0.68]{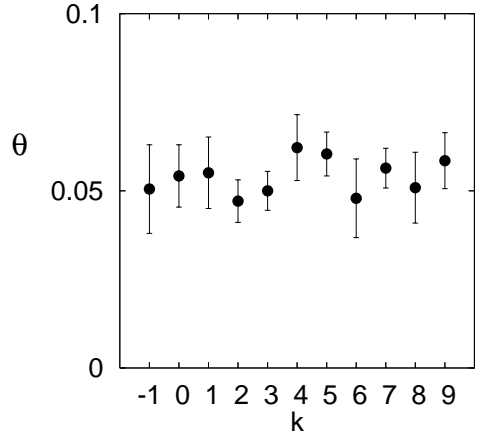}
\caption{
Value of $\theta$ for each of $k$
estimated by the 2nd order IDA of the combined quantity.    
}
\label{figure:r_vs_k}
\end{figure}

Thus we have evaluated $\theta$ using the combination of $m_0$,\ $m_2$ and $m_4$
as
\begin{equation}
  {m_0}^{\alpha}
  \left( 1+ c^{\prime}m_2 \right)^{\alpha^{\prime}}
  \left( 1+ c^{\prime\prime} m_4 \right) ^{\alpha^{\prime\prime}}\ \ \ 
        \label{eqn:combined}
\end{equation}
with 
\begin{equation}
  \alpha(2-\eta_0)+\alpha^{\prime}(4-\eta_0)+\alpha^{\prime\prime}(6-\eta_0)=0 \ \ \ 
\end{equation}
and
\begin{equation}
  \alpha+\alpha^{\prime}+\alpha^{\prime\prime}=1 \ \ \ 
\end{equation}
and $\eta_0=\frac{1}{4}$, 
which is also considered to behave like Eq.(\ref{eqn:test_function}) in general.
We can however anticipate that, by taking the combination of the three quantities,
the subleading logarithmic term may be cancelled
if we choose appropriate values for the parameters 
$\alpha$, $c^{\prime}$ and $c^{\prime\prime}$.
We have used the biased 2nd order IDA;
\begin{equation}
  \tau^2 Q_{m_2}(\beta ) f^{\prime\prime}(\beta )+ \tau Q_{m_1}(\beta ) f^{\prime}(\beta ) 
  + P_l(\beta ) f(\beta ) +R_k(\beta )=O( \beta^{m_2+m_1+l+k+3} )
\end{equation}
with $Q_{m_2}(0)=1$. The exponent $-2\sigma\theta$ can be
evaluated by the solution $\gamma$ of 
\begin{equation}
\gamma(\gamma-1)\frac{Q_{m_2}(\beta_c)}{{\beta_c}^2}
  -\gamma \frac{Q_{m_1}(\beta_c)}{\beta_c} 
    + P_l(\beta_c )=0.
\end{equation}
Here we adopt $\beta_c=1.1200$.
The 2nd order IDA can make more precise evaluation of $\theta$ than the 1st order IDA
for the function like Eq.(\ref{eqn:test_function}).
Fig.~7 is the plot of the estimation of $\theta$
by the 2nd order IDA for the 33rd order series 
of the test function (\ref{eqn:test_function}) 
with $\theta=\frac{1}{16}$, $\sigma=\frac{1}{2}$, and $A=0.3$.
Of course the 2nd order IDA also gives different values of $\theta$ for each of $k$
if $B\ne 0$, but if $B$ is small enough it can present precise estimation for the 
exponent of the leading singularity.  
By the analysis of the real combined quantity (\ref{eqn:combined}) 
we have found that the estimated values of $\theta$  
converge to $\theta=0.050(15)$ for all the range of $-1\le k\le 9$
in a domain of the set of parameters 
$-\alpha=3.6 - 3.0$, $c^{\prime}=2.8 - 4.0$ and $c^{\prime\prime}=2.8 - 4.0$.
In the analysis we have restricted $m_2+m_1+l+k+3=32$ with
$-1\le k\le 9$ and $m_2\ge 7$, $m_1\ge 7$, $l\ge 7$. 
The approximants that have near-by singularity 
(i.e. the zero of $Q_2(\beta)$ or $Q_1(\beta)$)
with $|\beta-\beta_c|/\beta_c<0.2$
have been excluded.
The most convergent result $\theta=0.054(10)$ is obtained for 
$\alpha=-3.143$, $c^{\prime}=3.134$ and $c^{\prime\prime}=3.139$.
The values of $\theta$ for each $k$ in this set of parameters are shown in Fig.~8.  
We note that this value $\theta=0.054(10)$ is  
consistent with $\theta=\frac{1}{16}$ 
predicted by the renormalization group.

\section{Summary}
  We calculated the high-temperature series for the zeroth moment 
(magnetic susceptibility) and
the second and fourth moments of the correlation function 
in the XY model on the square lattice
to order $\beta^{33}$ by using the improved algorithm 
of the finite lattice method.
The obtained long series have presented us an estimation for the value 
of the critical inverse temperature as $\beta_c=1.1200(1)$, 
which is consistent with 
the most precise value given previously by the Monte Carlo simulation. 
The critical exponent $\theta$ for the multiplicative logarithmic correction 
is evaluated using the combination 
of the three moments of the correlation function, giving $\theta=0.054(10)$, 
which is consistent with the value $\theta=\frac{1}{16}$ predicted 
by the renormalization group argument.

\begin{acknowledgments}
The author would like to thank P. Butera and M. Pernici for 
sending their high-temperature series before publication.
This work was supported in part by a Grant-in-Aid 
for Scientific Research (No.\ 16540353)
from the Ministry of Education, Culture, Sports, Science and Technology.
\end{acknowledgments}

\bibliography{correlation_function_r}

\end{document}